\documentclass{nature}

\usepackage{subfigure}
\usepackage{array}
\usepackage{float}

\usepackage{graphicx}
\usepackage{amsmath}
\usepackage{amssymb}
\usepackage{lineno}
\usepackage{caption}
\usepackage{booktabs}
\usepackage{multirow}
\usepackage{cite}
\usepackage{times}
\usepackage{color}
\usepackage{xurl}

\usepackage{docmute}
\usepackage[dvipsnames]{xcolor}

\newenvironment{changes}{\color{black}}

\title{\textbf{Switching exploration modes in human mobility}}

\author{Lu Zhong${}^{1,2,6}$, Lei Dong${}^{3,6}$, Qi Wang${}^{4}$, Chaoming Song${}^{5}$, Jianxi Gao${}^{1,2*}$}
\begin{document}

\maketitle

\begin{affiliations}
\item{Department of Computer Science, Rensselaer Polytechnic Institute, Troy, NY, USA}
\item{Network Science and Technology Center, Rensselaer Polytechnic Institute, Troy, NY, USA}
\item{Institute of Remote Sensing and Geographical Information Systems, School of Earth and Space Sciences, Peking University, Beijing, China}
\item{Department of Civil and Environmental Engineering, Northeastern University, Boston, MA, USA} 
\item{Department of Physics, University of Miami, Coral Gables, FL, USA}
\item{These authors contributed equally: Lu Zhong, Lei Dong.}\\
Email:gaoj8@rpi.edu
\end{affiliations}

\begin{abstract} 

Recent advances in human mobility research have revealed consistent pairwise characteristics in movement behavior, yet existing mobility models often overlook the spatial and topological structure of mobility networks. By analyzing millions of devices' anonymized cell phone trajectories, we uncover a distinct modular organization within these networks, demonstrating that movements within spatial modules differ significantly from those between modules. This finding challenges the conventional assumption of uniform mobility dynamics and underscores the influence of heterogeneous environments on human movement. Inspired by switching behaviors in animal movement patterns, we introduce a novel "switch mechanism" to differentiate movement modes, allowing our model to accurately reproduce both the modular structures of trajectory networks and spatial mobility patterns. Our results provide new insights into the dynamics of human mobility and its impact on network formation, with broad applications in traffic prediction, disease transmission modeling, and urban planning. Beyond advancing the theoretical and practical understanding of mobility networks, this work opens new avenues for understanding societal dynamics at large.


\end{abstract}

\section*{Introduction}

Understanding human daily mobility has been crucial for a multitude of applications\cite{barbosa2018human}, and over the past two decades, significant advancements have been made in uncovering shared regularities in human movement\cite{gonzalez2008understanding}. These discoveries encompass phenomena such as Lévy flights\cite{rhee2011levy,raichlen2014evidence}, the distribution of time allocation\cite{song2010limits}, visit frequencies\cite{schlapfer2021universal}, exploration tendencies and return regularities\cite{song2010modelling,pappalardo2015returners}, and inflation pattern\cite{zhong2024universal}, which have led to advances in human mobility models\cite{simini2012universal,barbosa2018human,alessandretti2018evidence,alessandretti2020scales,xu2021emergence}. However, when comparing the real data with the classical model, we find that the spatial and topological structure of the mobility network generated by the widely used human mobility model (EPR model) is very different from the real data (Fig. \ref{figure_1}a-d). Specifically, the stay points of users in real data are more spatially dispersed, bringing higher average shortest-path length and modularity in network metrics (Fig. \ref{figure_1}ef). At the mesoscale, the frequency of people's trips decreases with distance from home, but the decay rate is significantly lower than the model predictions (i.e., people make substantial trips far from home, Fig. \ref{figure_1}g). These inconsistencies suggest that our current understanding of human mobility is missing some important mechanisms.

As a mobile species, humans share many similarities with animals in movement patterns, and studies of animal mobility in recent years have found different mobility modes across natural areas\cite{benhamou2014scales,vilk2022ergodicity,nathan2022big,vilk2022unravelling}. For instance, due to resource patchiness, avians switch between a wide-ranging commuting mode of movement across patches and an intensive, area-restricted searching mode for prey within a local patch\cite{benhamou2014scales,vilk2022ergodicity}. The observed switch behavior inspires us to consider whether this mechanism also exists for human mobility and whether it happens to explain the inconsistency between the model and the data, as shown in Fig. \ref{figure_1}. In fact, cities share many similarities with natural areas that feature distinct clusters of resources (such as amenities and employment opportunities)\cite{batty2006hierarchy,louf2013modeling,sahasrabuddhe2021centre,cabrera2023inferring}. The mobility mode within and across spatial clusters of resources may be diverse.

To systematically test our hypothesis, we analyzed two large-scale cell phone datasets. {\changes{One dataset consists of six months of privacy-enhanced GPS trajectory data collected from two million anonymized users in the United States}}, while the other includes two weeks of call detail records from 300,000 anonymized users in Senegal(see Methods and Figs. S1). By constructing mobility networks from cell phone trajectories, we detect a polycentric modular structure of the network and find users exhibiting similar intra-module exploration modes but differing inter-module exploration modes. Leveraging these insights, we introduce a switch mechanism to differentiate the exploration mode.  This mechanism not only allows for precise replication of individual trajectories but also illustrates the emergence of polycentric structures within the trajectory networks. The introduction of the switch mechanism opens new avenues for predicting mobility flows, particularly in long-range movements, which is vital for accurate disease spread forecasting.

\section*{Results} 
\begin{figure*}[t!]
\centering
\includegraphics[scale=1.05]{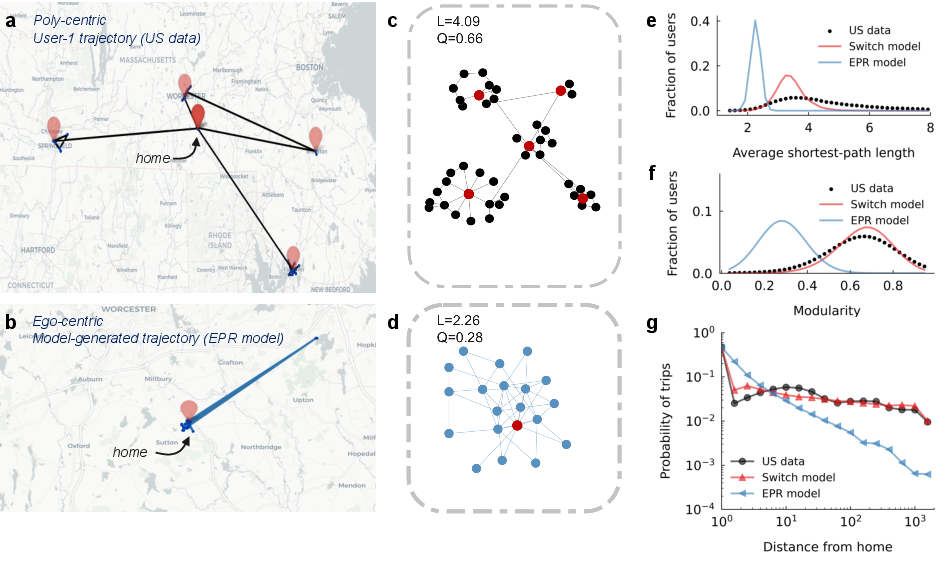}
\caption{\fontsize{10}{13.8}\linespread{1}\selectfont{\textbf{The polycentric modular structure of human mobility network.} \textbf{(a, b)} Trajectory of the anonymous cell phone user exhibits polycentricity, as opposed to the egocentricity of trajectory generated by a traditional model. The red marker denotes the center of activity. \textbf{(c, d)} Mobility networks associated with trajectories in \textbf{(a, b)}. The mobility networks are constructed regarding trajectory sequence and spatial distance, where nodes represent stay points and edges represent recorded travels. \textbf{(e)} The average shortest-path length distribution of users' mobility networks. \textbf{(f)} The geometric modularity distribution of users' mobility networks. \textbf{(g)} Probability of trips at a distance with the home as the reference point. The polycentric nature leads empirical mobility networks of high shortest-path length and high modularity and also have numerous travels distant from home. While the egocentric model, like the EPR model, fails to capture the network properties and generate distant travels. Our proposed model aligns with real-world mobility data and networks.}}
\label{figure_1}
\end{figure*}

\paragraph{Polycentric module structure in human mobility network.}
To characterize the structure of mobility data, we construct human trajectory networks based on cell phone data, where stay points as network nodes and paths connecting consecutive stay points are edges. The reciprocal of the spatial distance between stay points defines the edge weights. Hence, a smaller spatial distance results in a greater edge weight (see Methods). Figures \ref{figure_1}a and c display a trajectory of a real user and the corresponding mobility network, which has a polycentric structure with several activity regions. Comparing the empirical mobility network with the network generated by the classical exploration and preferential return (EPR) model, it is clear that the real-world mobility network manifests multiple modules. Conversely, the EPR model typically generates a single, highly connected module centered around the home location. Consequently, real-world mobility networks have significantly longer topological paths than the model, as shown in Fig. \ref{figure_1}e and Extended Data Fig. \ref{S_senegal_net}. In terms of network metrics, the EPR model typically exhibits a two-degree separation (the average shortest path length is two), implying that an individual can traverse any pair of locations through a single central transit hub. In contrast, real-world mobility networks' average shortest-path length follows a normal distribution with a median value of nearly four, implying a four-degree separation.

To quantify the observed polycentricity of the mobility network, we use geometric modularity ($Q$), which serves as a measure of the extent to which a network can be partitioned into distinct modules \cite{clauset2004finding}. Figure \ref{figure_1}f shows that the modularity of individual mobility networks exhibits a considerably high median value ${Q}$ around $0.66$. This is in contrast to the EPR model, which demonstrates lower modularity, with a median value of $0.28$. This higher value from empirical data suggests a high degree of segmentation of human mobility into multiple modules. In addition to the network metrics, the mesoscopic mobility patterns observed in the empirical data displayed in Fig. \ref{figure_1}g show notable differences with models, with significantly a broader tail for distant trips that are far from home. This distinct difference arises from the occurrence of long-range travels across modules, a feature conspicuously absent from previous models. See Extended Data Fig.\ref{S_senegal_net} for the results of Senegal mobility networks and Fig. S2-S3 for other details.

\begin{figure*}[t!]
\centering
\includegraphics[scale=1.05]{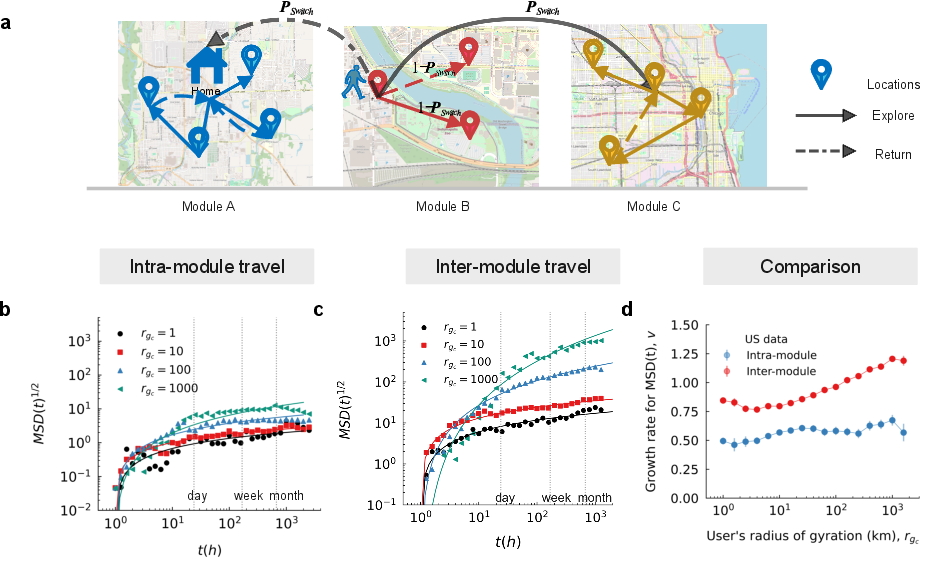}
\caption{\fontsize{10}{13.8}\linespread{1}\selectfont{\textbf{The switch  of exploration modes within and across modules.} \textbf{(a)} Illustration of the switch from intra-module travel mode to inter-module travel mode. \textbf{(b,c)} The time evolution of the mean squared displacement, $\text{MSD}(t)$, is used to quantify the spatiotemporal dynamics of inter- and intra-module mobility, where $\text{MSD}(t)^{1/2} \sim \text{log}(t)^v$. \textbf{(d)} The growth rate $v$ is plotted with error bars for users with different values of $R_{g_c}$. Despite variations in users' $R_{g_c}$, the analysis shows a lower growth rate $v$ for intra-module travels compared to inter-module travels, indicating a reduced tendency for exploration within modules. }}
\label{figure_2}
\end{figure*}

\paragraph{Switch of exploration mode.}  Our analysis of the mobility patterns reveals a polycentric modular structure, wherein locations within a module are spatially and topologically proximate, while intra-module travel requires significant displacement. We apply the Louvain method to extract modules from individual trajectory networks. The trajectory is then divided into two types of travel: intra-module travels, which involve movement within the local built environment, and inter-module travels, which refer to movement between different modules. After extracting the modules from the mobility network, as shown in Fig. \ref{figure_2}a, the user switches between intra-module and inter-module travels with switch tendency $P_{Switch}$.

To examine whether the dynamics of mobility behavior within modules differs with that across modules, we employ the mean square displacement (MSD). The MSD is defined as the squared displacement of an individual's position with regard to the reference position over time. Previous research suggests $\text{MSD}(t)^{1/2} \sim \log(t)^v$, indicating subdiffusive dynamics and a strong tendency to revisit familiar places \cite{song2010modelling}. A lower value of $v$ indicates a stronger propensity to return (smaller tendency to explore). As depicted in Fig. \ref{figure_2}b-d, the growth rate $v$ for intra-module movement differs from that for inter-module movement, challenging prior studies that assume uniform $v$ across all spatial scales. To determine if the growth rate $v$ is consistent among individuals, we classify users by their radius of gyration $R_{gc}$. Figure \ref{figure_2}d illustrates that, while the growth rate $v$ for intra-module movements remains relatively stable, the rate for inter-module movements increases for users with a larger $R_{gc}$. See Fig. S4 for the results for Senegal data.

To calibrate the switch in mobility behavior, we measure the tendencies for exploration and return, $P_{w}$ and $1-P_{w}$ for within-module movements, and $P_{c}$ and $1-P_{c}$ for between-module movements, respectively. The exploration tendency $P_{w}$ ($P_{c}$) is calculated as the proportion of unique locations (module) visited relative to the total movements within a module (cross modules). Figures S4-S5 demonstrate that $P_{w} \sim S_w^{-\gamma_w}$ and $P_{c} \sim S_c^{-\gamma_c}$, whereas the $\gamma_w$ consistent across users with different radii of gyration ($R_{g_c}$), but $\gamma_c$ decreases for users with higher radii of gyration. It indicates that the tendency to explore within modules is similar across users, but the tendency to explore across modules increases with larger $R_{g_c}$, same as demonstrated in Fig.  \ref{figure_2}e.

\paragraph{Predictions of switch mechanism.} Building on the identified behavior, we introduce the switch mechanism and the calibrated parameters into the standard exploration/preferential return scheme model \cite{song2010modelling} (see Methods and Fig. S6-S8). As shown in Fig. \ref{figure_1}, our switch model accurately depicts the characteristics of mobility networks. The model’s projection of the average path length aligns closely with empirical data, with a median value of approximately ${L}_{\text{Switch}}=3.78$ (Fig. \ref{figure_1}e). Moreover, our model captures the modular structure observed in empirical mobility networks. As shown in Fig. \ref{figure_1} and Extended Data Fig. \ref{S_senegal_net}, our model yields a high modularity measure, closely matching empirical findings with $ {Q}_{\text{Switch}}=0.68$ for U.S. data and $ {Q}_{\text{Switch}}=0.35$ for Senegal data. This stands in sharp contrast to the smaller modularity offered by the EPR model, with $ {Q}_{\text{EPR}}=0.28$ for U.S. data and $ {Q}_{\text{EPR}}=0.20$ for Senegal data. For the clustering coefficient, indicative of a propensity for triangle paths,  our model results align with empirical values ($ {C}_{\text{Switch}}=0.31$ for U.S. data and $ {C}_{\text{Switch}}=0.32$ for Senegal data). The EPR model, in contrast, predicts higher clustering coefficients ($ {C}_{\text{EPR}}=0.38$ for U.S. data and $ {C}_{\text{EPR}}=0.53$ for Senegal data).

\begin{figure*}[t!]
\centering
\includegraphics[scale=1.2]{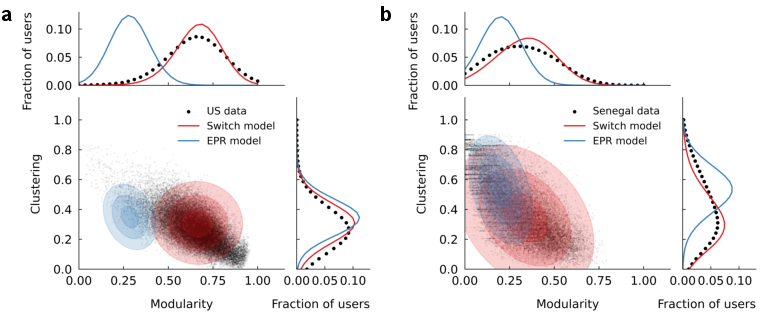}
\caption{\fontsize{10}{13.8}\linespread{1}\selectfont{\textbf{Switch mechanism depicts the modular structures in mobility networks.} \textbf{(a, b)} The unusual inverse relationship between modularity and the clustering coefficient of mobility networks in U.S. data \textbf{(a)} and Senegal data \textbf{(b)}. The polycentric switch model captures the inverse relationship, effectively generating a network structure with spatially separated modules but a limited number of triangle paths, aligning with empirical data. }}
\label{figure_3}
\end{figure*}

Contrary to conventional network science beliefs, where high modularity typically results in a larger clustering coefficient and vice versa\cite{newman2006modularity}, we observe an unusual negative correlation between modularity and clustering coefficient in mobility networks. This observation uncovers a key aspect of the polycentric nature of human mobility networks, further underscoring the validity of our model. Such a negative correlation embodies the polycentric modular structure of human mobility networks, where each module has a star-like structure dominated by a few transit hubs. These hubs route most paths, leaving few opportunities for triangle formation, thereby, resulting in a low clustering coefficient. At the same time, the high degree of module separation results in high modularity. The distinct contrast between our model and the traditional egocentric EPR model becomes particularly apparent in the scatter plot of modularity versus clustering coefficient, as depicted in Fig. \ref{figure_3}. Our proposed model and the empirical data it represents occupy the bottom-right corner of the plot, symbolizing a combination of high modularity and low clustering coefficient. In stark contrast, the egocentric EPR model, represented by a single, highly interconnected module, occupies the top-left corner, corresponding to low modularity and a high clustering coefficient.

\begin{figure*}[t!]
\centering
\includegraphics[scale=1.2]{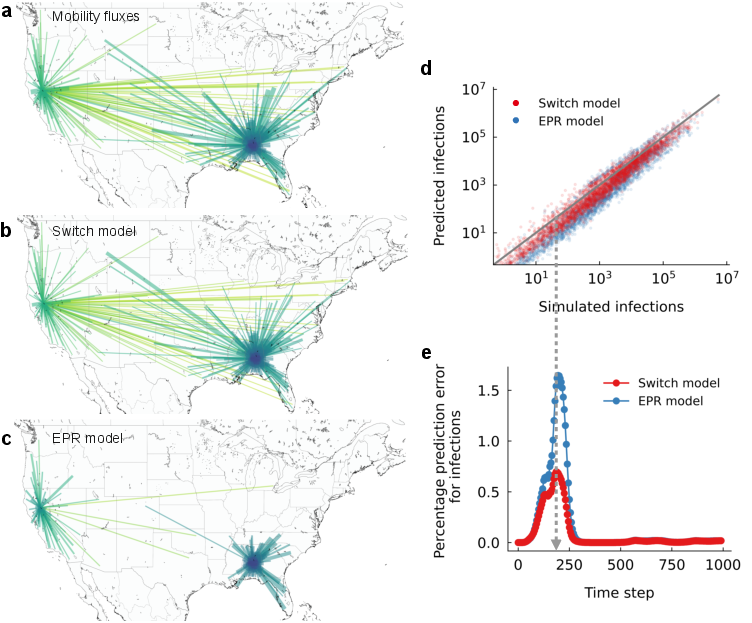}
\caption{\fontsize{10}{13.8}\linespread{1}\selectfont{\textbf{Switch mechanism predicts long-range travels and infections.} \textbf{(a, b, c)} The mobility flux originates from two counties (located in Alabama and California, respectively). The switch model demonstrates superior alignment with data, particularly in generating long-range fluxes. While the egocentric EPR model generates fewer long-range fluxes. To evaluate the models' efficacy in predicting infections, we simulate the spread of disease on predicted and empirical cross-county fluxes, with an initial outbreak in Alabama County. \textbf{(d)} The comparison between model-predicted infections and simulated infections in all counties at the snapshot time $t=150$. \textbf{(e)} Models' prediction error at the entire time course. The switch model demonstrates low prediction error, indicating its proficiency in accurately predicting infections. }}
\label{figure_4}
\end{figure*}

Furthermore, our model better predicts long-range movements. Figure \ref{figure_1}g shows that our model fits with empirical data in capturing travels that are distant from home. After integrating all users' travels, we show that our model also exhibits superior agreement with empirical data in county-level mobility fluxes, particularly in long-range mobility trips, as illustrated in Fig. \ref{figure_4}a-c and  Extended Data Fig. \ref{S_collective_travels}. The polycentric nature of the mobility network generated by our model enables a direct connection between two distant counties. In contrast, the egocentric EPR model tends to overestimate short-range fluxes, thereby failing to reproduce direct long-range connections. On the predicted and empirical mobility fluxes, we simulate the disease spread using a meta-population model (see Methods for details of the simulation). Our model excels in predicting the spread of disease, while the EPR model overestimates local infections and underestimates infections distant from sources. This discrepancy is evident in the example comparison at the same time shown in Fig. \ref{figure_4}d. When evaluating the overall progression of infections over time in Fig. \ref{figure_4}e, our model consistently outperforms the EPR model with prediction errors up to four times lower.

\section*{Discussion} 

Our study shows the nuanced dynamics of human mobility, particularly identifying the switch of exploration modes across spatial scales and diverse populations. The underlying mechanisms contribute to the uniqueness of the human trajectory network, characterized by high modularity and clustering, leading to a polycentric nature where each module contains a hub. By categorizing users according to various demographic factors, including age, gender, race, ethnicity, poverty level, and household income (Extended Data Fig. \ref{S_demographic_net}), we found that the modular structure and the topological features of the trajectory network remained consistent regardless of variations in individual demographics. 

The switch mechanism highlights the impact of heterogeneous distributions and clustered resources in urban environments on human mobility\cite{louf2013modeling,sahasrabuddhe2021centre,cabrera2023inferring}. It reveals how humans switch between long-range, cross-module travel modes and short-range, within-module travel modes to meet daily human needs such as food, social interactions, and other necessities. The model's ability to accurately predict both individual and collective mobility patterns presents significant potential for future practical applications, particularly in urban planning \cite{frank2001built}, epidemic prevention \cite{belik2011natural,chang2021mobility}, and mitigating activity inequality \cite{althoff2017large,wang2018urban,athey2021estimating}.

Moreover, the switch mechanism aligns with fundamental principles observed in animal and molecular movement \cite{benhamou2014scales,golan2017resolving,vilk2022ergodicity}, as well as key concepts in movement ecology \cite{meekan2017ecology,nathan2022big}. This cross-disciplinary connection broadens the scope of our model, suggesting its potential utility across diverse research domains. Future efforts could focus on refining the model to address specific scenarios or integrating it with complementary approaches to enhance its applicability. These insights not only highlight the importance of understanding mobility patterns at multiple scales but also open avenues for collaborative studies across disciplines to tackle pressing challenges in movement dynamics and beyond.

\section*{Methods}
\subsection{Data.} We analyzed human movement using two datasets: the U.S. dataset from Cuebiq Inc. and the Senegal dataset from the D4D Senegal Challenge\cite{de2014d4d,schlapfer2021universal}. {\changes{The U.S. dataset contains anonymized location data of 42 million devices from January to June 2020. All devices opted into anonymized data collection for research purposes. In addition to de-identifying the data, the data provider also obfuscates home locations to the census block group level to preserve privacy.}} Using the Infostop algorithm\cite{aslak2020infostop}, we processed this data to identify stay points, focusing on 2.1 million users with over thirty days of data. We divided the dataset based on the pre and post-lockdown periods starting March 11, 2020, mainly using the pre-lockdown data for analysis and validation with the post-lockdown segment. The Senegal dataset comprises anonymized call records from 2013, segmented into 25 two-week periods with about 44 million records from 300,000 users, capturing movement linked to mobile towers. Table. S1 summarizes the basic statistics of the two datasets. 

For the data preprocessing,  we categorize the geographical locations of all users by employing the H3 indexing system at a resolution of 12. This system divides the physical space into hexagonal cells, each with an edge length of roughly 9 meters. We determine users' home locations based on the cell they visit most frequently during the night, specifically between 8 p.m. and 8 a.m.

\subsection{Characterizing human trajectory networks.} For each user's sequence of stay points $T=\{\theta_{1},...,\theta_{i}...\}$ where $i$ indexes the sequence, we construct the trajectory network $G(T)$. In this network, nodes represent stay points, and edges correspond to consecutive travels between these points. To ensure that all weights are non-negative and that shorter distances between points yield larger weights, we define the edge weight between two consecutive points $(\theta_{i},\theta_{i+1})$ as $w(\theta_{i},\theta_{i+1})=\log(\frac{\hat{d}}{d(\theta_{i},\theta_{i+1})})$, where $d(\theta_i, \theta_{i+1})$ represents the spatial distance between consecutive stay points, and $\hat{d}$ is a predefined maximum jump distance. This inverse distance weighting reflects Tobler’s first law of geography\cite{miller2004tobler}, suggesting that closer entities have stronger spatial interactions. We set $\hat{d}$ as $4,000$km for the U.S. and $1,000$km for Senegal.

To analyze the trajectory network, we calculate key metrics\cite{barabasi2014network} such as the shortest path length ($L$), average clustering coefficient ($C$), and weighted modularity\cite{newman2006modularity} ($Q$). For the detection of modules within the weighted directed trajectory network $G(T)$, we employ the Louvain method\cite{fortunato2010community}. This method helps identify sub-networks, or modules, composed of stay points that are closely linked both spatially and topologically. 

\subsection{Characterizing mobility dynamics within and across modules.} Building on existing studies, we characterize mobility dynamics within and between modules by calculating the mean square displacement (MSD). MSD measures the squared deviation of an individual's position from a reference position, averaged across various movement paths, defined as $\text{MSD}(t) = \langle \Delta x^2(t) \rangle$. If the movement is superdiffusive, then $\langle \Delta x^2(t) \rangle \sim t^v$ with $v>1$. In the case of Brownian motion, $\langle \Delta x^2(t) \rangle \sim t^v$ with $v=1$. However, for subdiffusive dynamics, $v<1$. For both intra-module and inter-module mobility behaviors, we find that:
\begin{equation} 
\text{MSD}(t)^{1/2} \sim \log(t)^v 
\end{equation}
This relationship indicates that MSD follows a growth slower than logarithmic, suggesting an ultraslow diffusive process. Moreover, the growth rates $v$ vary between intra-module and inter-module behaviors. Larger growth rates $v$ suggest a higher (lower) likelihood of exploring new locations (returning to familiar locations).

\subsection{Switch mechanism.} We introduce the switch mechanism to the typical individual mobility model, i.e., the EPR 
 model\cite{song2010modelling}, to account for the difference in intra- and inter-module mobility behaviors. As illustrated in Fig. S6, users with a unique radius of gyration $R_{g_c}$ will initiate their first move from their home locations. After a waiting time $\Delta t$, we assume that the user who is at module $i$ has the probability of $P_{Switch}$ of switching inter-module mode or $1-P_{Switch}$ to continue stay in current module $i$ :

Option (1): Inter-module mode. The individual may either transition to a new module $j$ with a probability $P_c$ or return to a frequently visited module with a probability $1-P_c$. The count of inter-module movements $n_{c}$ increases to $n_{c}+1$. Should they venture into a new module, the count of unique modules visited, $S_c$, will also increase by one to $S_c+1$. 

Option (2): Intra-module mode. The individual has a probability $P_{w_i}$ of exploring a new location within the current module $i$  or a probability $1-P_{w_i}$ of returning to previously visited, familiar locations. The count of movements $n_{w_i}$ increases to $n_{w_i}+1$. If a new location is explored, the count of unique locations within the module, $S_{w_i}$, will increment by one to $S_{w_i}+1$. The exploration ends when $S_{w_i}$ reaches the criteria of unique locations.

To validate our proposed model, we first need to estimate the parameters from empirical data. Besides the distributions of the inter-module radii of gyration $P(R_{g_c}) \sim R_{g_c}^{-(1+\eta)}$ and distributions of waiting for time $P(\Delta t)$ and jump distance $P(\Delta r)$ (see Supplementary Information and Figs. S1), we calibrate the probability set  $P_{w}$, $P_c$, and $P_{Switch}$. The probability $P_{w}$ (and $P_c$) pertaining to the inclination for intra-module (inter-module) exploration is evaluated by observing changes in the number of distinct intra-module locations $S_w$ over a span of movements $n_w$. As depicted in Fig.S4-S5, the relationship between $P_w$ and $\Delta S_w$ can be approximated as $P_w \sim \Delta S_w = \rho_w S_w^{-\gamma_w}$. Similarly, $P$ and $\Delta S$ exhibit a relationship of $P_c \sim \Delta S = \rho S^{-\gamma_c}$. It's notable that while parameter $\gamma_w$ is nearly the same for individuals with different $R_{g_c}$,  parameter $\gamma_c$ decreases for individuals with higher $R_{g_c}$. When we compare the $\gamma_w$ and $\gamma_c$ with $R_{g_c}$ as the reference axis in Fig.\ref{figure_2}h,  it becomes apparent that $\gamma_w $ is approximately constant, while $\gamma_c$ follows a logarithmic trend, specifically $\gamma_c \sim -\log(R_{g_c})$.  The parameters $\rho_w$ and  $\rho_c$ are respectively estimated around 0.6 and 1, as demonstrated in Fig. S7. Furthermore, the probability $P_{Switch}$, which represents the tendency to transition from the current module to another module, is set to 0.14.

In cases where $P_{Switch}=0$, the model simplifies to the EPR model, where all individuals only take intra-module travels such that all individuals share the same exploration tendency. See Supplementary Information and Fig. S9 for sensitivity analysis of parameters affecting model results.

\subsection{Estimating diseases spread through predicted mobility patterns.} By integrating the trajectories of all users, we can analyze the empirical cross-county fluxes and compare them with predictions made by individual mobility models. To simulate the spread of disease based on these cross-county mobility fluxes, we utilize the meta-population susceptible–infected–recovered  model\cite{brockmann2013hidden}. In this model, the disease evolution within each county is governed by three states: susceptible, infectious, and removed, and the transmission of the disease between counties is described by the mobility fluxes $T_{ij}$ from county $j$ to county $i$. We select Alabama's Augusta County as the initial outbreak location and introduce a single infection, simulating the disease's spread over a time span of 1000 days with a reproductive number of $4$, an infection rate of 0.4 day$^{-1}$, and cross-county migrate rate of 0.004 day$^{-1}$. To ensure a fair comparison, we normalize the model-predicted mobility fluxes in order to maintain the same total mobility fluxes as observed in the empirical data. The accuracy of predicting infections between empirical and model-predicted mobility fluxes is assessed using the Mean Absolute Percentage Error at each time step.

\clearpage
\section*{References}

\clearpage
\begin{addendum}

\item[Acknowledgments]We thank Jinzhu Yu for his assistance with pre-processing the Senegal mobility data and fruitful discussion. J.G. and L.Z. acknowledge the support of the US National Science Foundation under Grant No. 2047488. L.D. was supported by the National Natural Science Foundation of China under Grant No. 42422110 and the Fundamental Research Funds for the Central Universities, Peking University.

\item[Author contributions] L.Z., L.D., Q.W., and J.G. conceived the project and designed the experiments; Q.W. collected and analyzed the US data; L.Z. analyzed the Senegal data; L.Z., L.D., J.G., and C.S. carried out theoretical calculations and performed the experiments; all authors wrote and edited the manuscript.

\item[Competing interests] The authors declare no competing interests.

\item[Correspondence and requests for materials] should be addressed to J.G.

\item[Additional information]
Supplementary Notes 1-4, including Supplementary Figures 1-11, and Supplementary Tables 1-3.

\item[Data and code availability] Data files and the Python script have been deposited in \url{https://github.com/lucinezhong/Switch_Mechanism_Human_Mobility}

\end{addendum}
\newpage
\clearpage

\newpage

\clearpage
\newpage

\clearpage

\clearpage
\section*{Extended Data Figures}
\renewcommand\thefigure{\arabic{figure}}
\setcounter{figure}{0}

\begin{figure*}[h!]
\renewcommand\figurename{\bf{Extended Data Fig.}}
\centering
\includegraphics[scale=1.2]{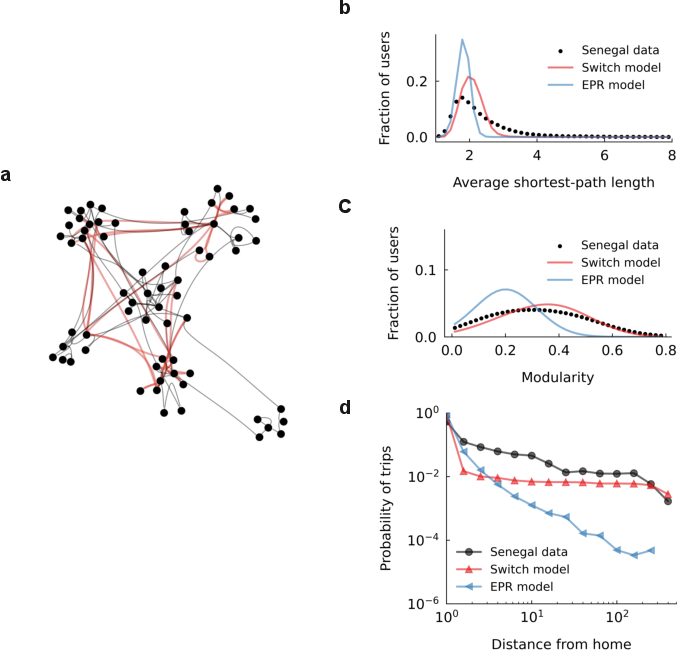}
\caption{\fontsize{10}{13.8}\linespread{1}\selectfont{\textbf{Mobility network ssociated with trajectories for Senegal data.}} \textbf{(a)} Visualizations
of example trajectory networks. Edges are in black but triangle links are in red.  \textbf{(b)} The average shortest-path length distribution of users' mobility networks. \textbf{(c)} The geometric modularity distribution of users' mobility networks. \textbf{(d)} Probability of trips at a distance with the home as the reference point. Same as Fig.\ref{figure_1}, the egocentric EPR model fails to fit with real-world data. The distribution of real-world lengths of the shortest path shows a long-tail pattern, and the extent of geometric modularity is more significant.}
\label{S_senegal_net}
\end{figure*}

\begin{figure*}[h!]
\renewcommand\figurename{\bf{Extended Data Fig.}}
\centering
\includegraphics[scale=1.0]{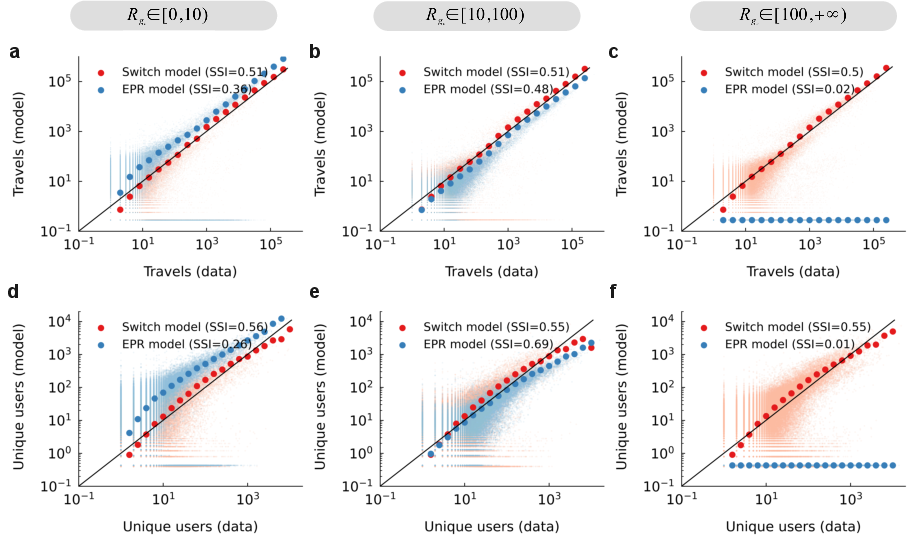}
\caption{\fontsize{10}{13.8}\linespread{1}\selectfont{\textbf{The switch mechanism better depicts collective mobility.}} Model-predicted and empirical data of mobility fluxes (\textbf{a, b, c}) and unique users (\textbf{d, e, f}) between counties are compared in three user groups with $R_{g_c}$ in ranges $[[0, 10); [10, 100); [100, \infty)]$. A higher SSI means a better match with empirical data. The egocentric EPR model overestimates the travel for user groups in 
$R_{g_c} \in [0,10]$ that primarily have small-distance travels.  The polycentric switch model fits better with data with higher SSI for all user groups. }
\label{S_collective_travels}
\end{figure*}

\begin{figure*}[h!]
\renewcommand\figurename{\bf{Extended Data Fig.}}
\centering
\includegraphics[scale=0.9]{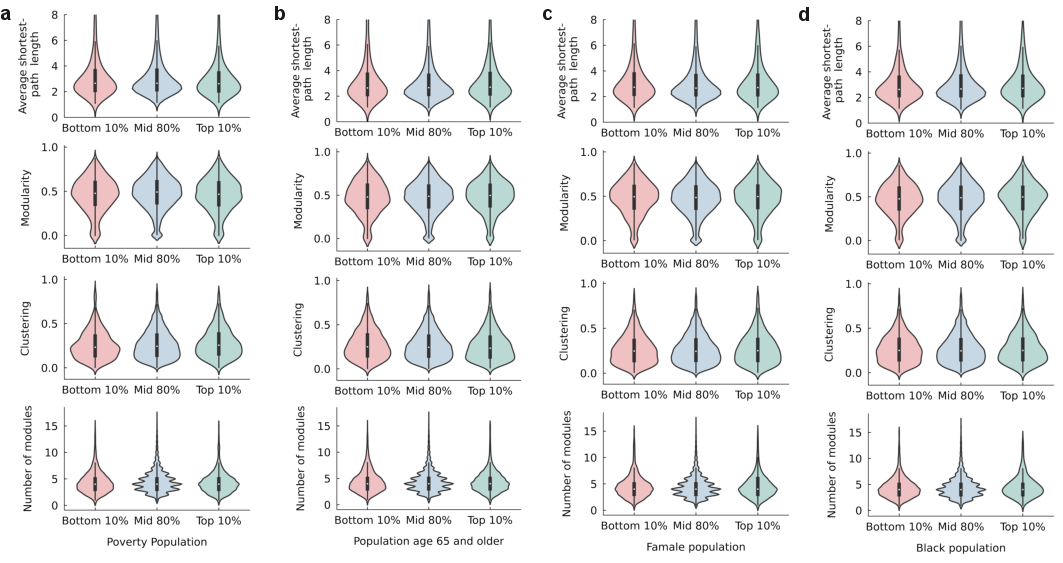}
\caption{\fontsize{10}{13.8}\linespread{1}\selectfont{\textbf{Mobility network characteristics across diverse populations in different demographic attributes.}}  When categorizing users based on the proportions of the poverty population in their home locations \textbf{(a)}, the elderly population (age 65 and older) \textbf{(b)}, the female population \textbf{(c)}, and the black population \textbf{(d)}, the distribution of average shortest path length, clustering, modularity, and module numbers remain consistent across user groups. The median values are indicated in white.}
\label{S_demographic_net}
\end{figure*}

\end{document}


\baselineskip24pt

\maketitle

\tableofcontents

\newpage

\section*{Supplementary Note 1: Related Work}
\addcontentsline{toc}{section}{\protect\numberline{}Supplementary Note 1: Related Work}%

\subsection*{1.1 MSD and subdiffusion}
\addcontentsline{toc}{subsection}{\protect\numberline{1.1} MSD and subdiffusion}%

To characterize human movement, researchers use 
the Mean Squared Displacement (MSD) to measure the deviation of position with regard to the reference position over time. The MSD is defined as follows,
\begin{equation}
    \text{MSD}(t)=\langle \Delta x^{2}(t)\rangle =\langle (x(t)-x_0(t))^2\rangle
\end{equation}
If the MSD follows $\langle  \Delta x^{2}(t)\rangle \sim t^{v}$ with $v=1$, it is a classical Brownian motion \cite{einstein1905movement,meroz2010subdiffusion}. When $v<1$, it is described as subdiffusion; where $v>1$, it is described as superdiffusive. According to the literature \cite{kapanidis2018understanding,song2010modelling,rhee2011levy}, we know that human motions are subdiffusion processes. Specifically, human mobility is a stronger subdiffusion process whose MSD follows a slower than logarithm growth, i.e., $\langle \Delta x^{2}(t) \rangle^{1/2} \sim (\log (t))^{v}$.

\subsection*{1.2 EPR model}
\addcontentsline{toc}{subsection}{\protect\numberline{1.3} EPR model}%

 To account for the nature of humans returning to familiar locations on a daily basis, the EPR model\cite{song2010modelling,pappalardo2016human} incorporates two mechanisms: exploration and preferential return. The exploration mechanism specifies that an individual will explore a new location with probability $P=\rho S^{-\gamma}$, and the preferential return mechanism specifies that an individual will return with probability $1-P$ to the historically visited location $i$ according to locations' past visitation frequency $\Pi_I= f_i$. $S$ is defined as the number of distinct locations precisely visited. Two parameters, $\rho$ and $\gamma$ ($\rho=0.6$ and $\gamma=0.21$) determine the exploration tendencies of individuals. 

\section*{Supplementary Note 2: Empirical results}
\addcontentsline{toc}{section}{\protect\numberline{}Supplementary Note 2: Empirical results}%

\subsection*{2.1 Mobility properties} 
\addcontentsline{toc}{subsection}{\protect\numberline{2.1} Mobility properties}%

\begin{figure*}[t!]
\centering
\includegraphics[scale=1.0]{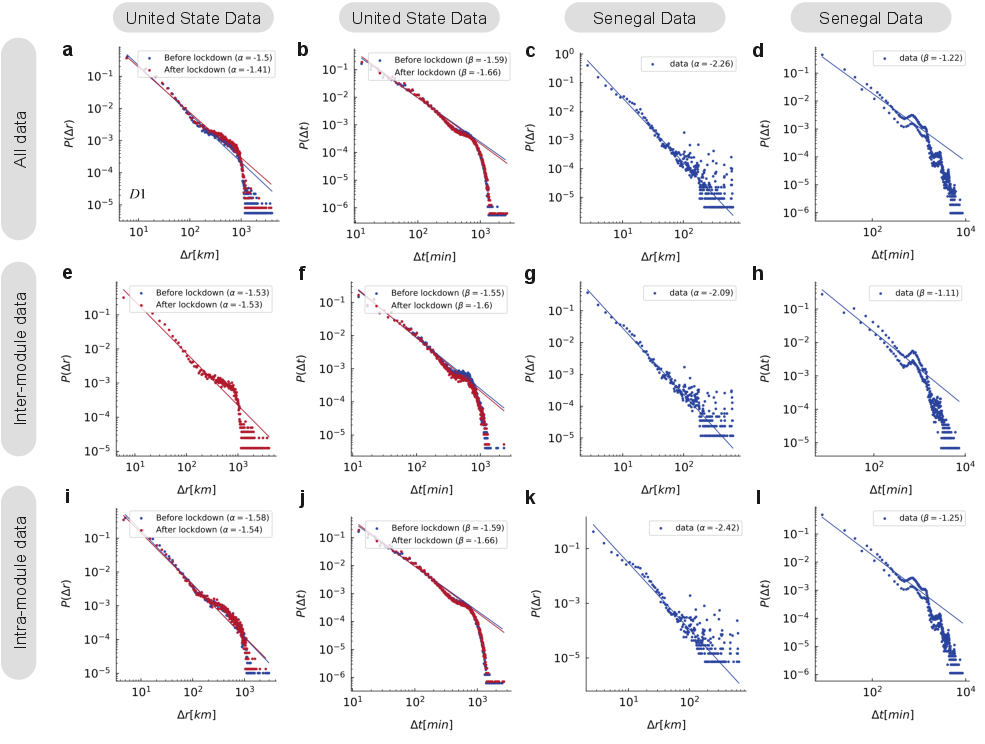}
\caption{\bf{Jump-distance distribution $P(\Delta r)$ and stay-time distribution $P(\Delta t)$ for U.S. data and Senegal data.}
\label{SI_travel_distance}}
\end{figure*}

We use the fundamental measures of jump lengths and waiting time to characterize human mobility. Fig. \ref{SI_travel_distance} shows the jump-distance $\Delta r$ and waiting-time $\Delta t$ distributions for the United States data and Senegal data. These distributions are approximated by $P(\Delta r) \sim \Delta r^{-(1+\alpha)}$ and $P(\Delta t) \sim \Delta r^{-(1+\beta)}$ where $\alpha=0.50$ and $\beta=0.60$ for United State data; and $\alpha=1.26$ and $\beta=0.22$ for Senegal data. There are no significant differences between intra-module travels and inter-module travels.

\subsection*{2.2 Mobility network properties} 
\addcontentsline{toc}{subsection}{\protect\numberline{2.2} Mobility network properties}%

Human mobility trajectory is often characterized by a network where nodes represent the visited locations and directed edges represent travels between them. Existing research has been based on network features such as motifs to model human movements and makes predictions based on them\cite{schneider2013unravelling}. But they either primarily focus on collective flow in transportation networks or discard spatial information of locations \cite{hossmann2011complex,schneider2013unravelling}. As stated in the Methods, we construct human mobility trajectory as a network where the edge weights represent spatial distance. We use three typical measures, average shortest-path length, clustering coefficient, and modularity to depict the trajectory network. As shown in Fig. 1 and Extended Data Fig. 1 the modularity is normally distributed with median value ${Q}_{Senegal}=0.31$  for Senegal data. The average shortest path length is also normally distributed with median value ${L}_{Senegal}=2.07$. Compared with the U.S. data in the manuscript, the Senegal data has a smaller modularity value and a smaller average shortest-path length. One possible reason is that Senegal data is based on Call Detail Records, which is more coarse-grained leading to missing explicit intra-module travels. Similar to the results in the manuscript, the trajectory networks generated by the EPR model cannot capture the network characteristics of human mobility trajectories.  In Extended Data Fig. 1b-c, the distributions of modularity and shortest path of EPR-model-generated trajectory networks deviate from that empirical Senegal data with ${Q}_{EPR}=0.20$ and ${L}_{EPR}=1.81$. Our switch model surpasses the EPR model in effectively capturing the long-tail distribution of shortest-path lengths observed in empirical data. Summarily, both results in the U.S. data and Senegal data present the high modularity and low clustering coefficient of human trajectory networks and existing models cannot capture the characteristics.

\paragraph{Count of unique locations.} As shown in Fig. \ref{SI_module_size}, the module size (number of unique locations within modules) at varying distances from home remains consistent. For the U.S. datasets, this value stabilizes at approximately 10, while for the Senegal dataset, it remains around 5.

\begin{figure*}[t!]
\centering
\includegraphics[scale=1.2]{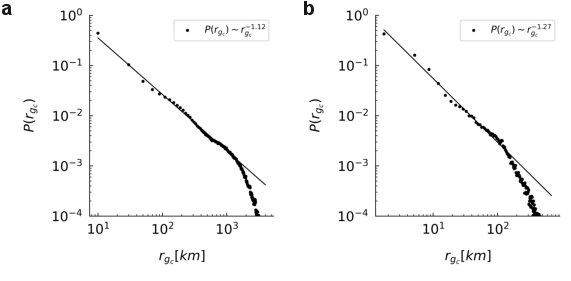}
\caption{{\bf Users’ inter-module radius of gyration distribution $P(R_{g_c})\sim R_{g_c}^{-(1+\eta)}$  for U.S. data and Senegal data.} The data presents a power-law decay with $\eta=0.12$ for the United States
and $\eta=0.27$ for Senegal.}
\label{SI_Rgc}
\end{figure*}

\begin{figure*}[t!]
\centering
\includegraphics[scale=1.1]{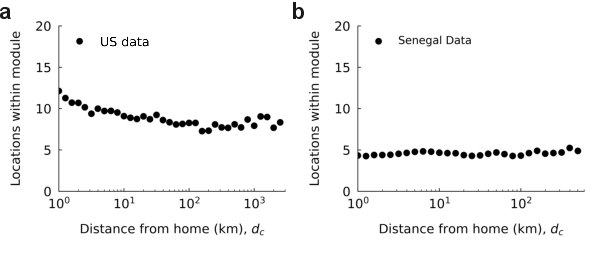}
\caption{\bf{ Number of unique locations within modules for U.S. data and Senegal data. }}
\label{SI_module_size}
\end{figure*}

\paragraph{Inter-module radius of gyration.} Individuals are different in the way that modules are distributed. To measure how the modules are
spatially distributed for each individual, we use the radius of gyration at the inter-module level, denoted as 
\begin{equation}
    R_{g_c}=\sqrt{\frac{1}{N} \sum_i^N d_{c_i}^2}
\end{equation}
where $N$ is the number of modules and $d_{c_i}$ is the distance between the centroid of module $i$ and the home location. The smaller $R_{g_c}$, users' modules are overlapped near the home location. The larger $R_{g_c}$, modules are more distant from each other. We analyzed millions of user trajectories in the United States and Senegal data in Fig. \ref{SI_Rgc},  the distribution of users’ $R_{g_c}$ represents a power-law decay $P(R_{g_c}) \sim R_{g_c}^{-(1+\eta)}$ with $\eta=0.12$ for U.S. Data and $\eta=0.27$ for Senegal Data.

\section*{Supplementary Note 3: Switch mechanism}
\addcontentsline{toc}{section}{\protect\numberline{}Supplementary Note 3: Switch mechanism}%

\subsection*{3.1 Model}  
\addcontentsline{toc}{subsection}{\protect\numberline{3.1} Model}%

\begin{figure*}[t!]
\centering
\includegraphics[scale=1.0]{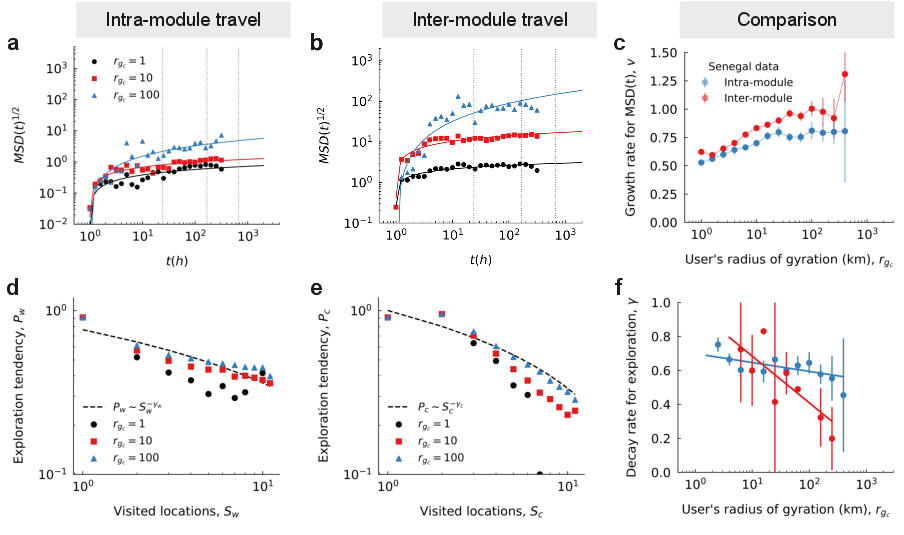}
\caption{{\bf Switch of exploration modes within and across modules in Senegal data.} \textbf{(a,b)} The time evolution of the mean squared displacement, $\text{MSD}(t)$, is used to quantify the spatiotemporal dynamics of inter- and intra-module mobility, where $\text{MSD}(t)^{1/2} \sim \text{log}(t)^v$. \textbf{(c)} The growth rate $v$ is plotted with error bars for users with different values of $R_{g_c}$. \textbf{(d,e)}  Intra-module exploration tendency follows a power-law relation, $P_w \sim S_w^{\gamma_w}$, and inter-module exploration tendency follows $P_c \sim S_c^{\gamma_c}$. \textbf{(f)} Comparison of exploration tendency $\gamma_w$ and $\gamma_c$. }
\label{SI_Senegal_MSD}
\end{figure*}

\begin{figure*}[t!]
\centering
\includegraphics[scale=1.0]{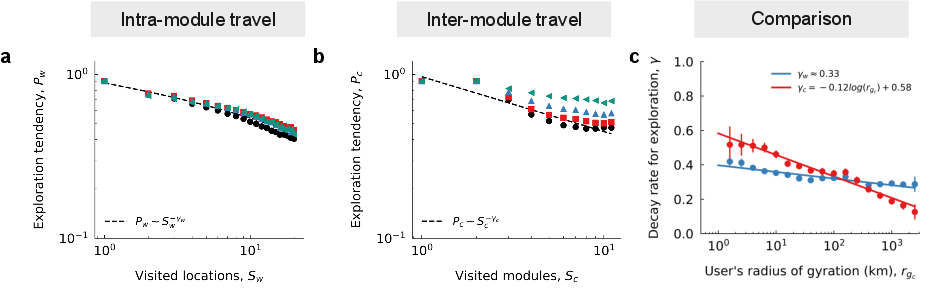}
\caption{{\bf Switch of exploration modes within and across modules in US data.} \textbf{(a,b)}  Intra-module exploration tendency follows a power-law relation, $P_w \sim S_w^{\gamma_w}$, and inter-module exploration tendency follows $P_c \sim S_c^{\gamma_c}$. \textbf{(c)} Comparison of exploration tendency $\gamma_w$ and $\gamma_c$. }
\label{SI_US_MSD}
\end{figure*}

To compare the difference between human mobility behavior within and cross modules, we show the mean square displacement ($\text{MSD}_w(t)$ and $\text{MSD}_c(t)$) and exploration tendencies for the intra-and inter-module travels in both US dataset and Senegal dataset (see Fig.\ref{SI_Senegal_MSD} and Fig.\ref{SI_US_MSD}). It is evident that users with different $R_{g_c}$ have different inter-module behavior patterns while having the same intra-module behavior pattern on the basis of the measures.  In Fig. \ref{SI_Senegal_MSD}, $\text{MSD}_c(t)^{1/2} \sim (\log(t))^{v_c}$ and user groups with smaller $R_{g_c}$ have a lower saturation value of $\text{MSD}_c(t)$. To clearly show the difference, we set users $R_{g_c}$ as independent variables and the growth rate $v_w$ and $v_c$ as the dependent variables. New models considering intra- and inter-module mobility differences are needed. 

\begin{figure*}[t!]
\centering
\includegraphics[scale=0.97]{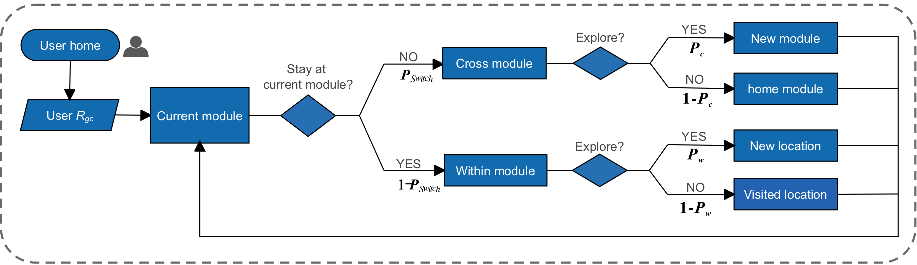}
\caption{{\bf Schematic description of switch mechanism.} After waiting for time $\Delta t$, an individual with $R_{g_c}$ at the current module can either travel to another module with probability $P_{Switch}$ or decide to keep staying within the current module with probability $1-P_{Switch}$. If staying at (jumping out the) current module, the individual will visit new locations (new module) with probability $P_w$ ($P_c$) or return previously visited locations within the current module (return home module). }
\label{SI_Model}
\end{figure*}

\begin{figure*}[t!]
\centering
\includegraphics[scale=1.0]{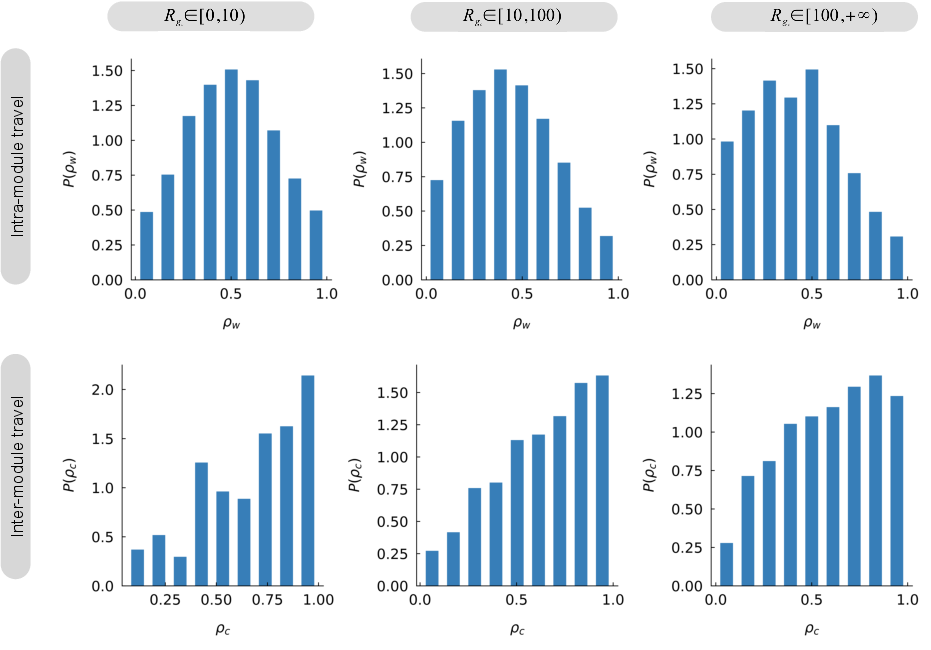}
\caption{{\bf The parameter $\rho_w$ and $\rho_c$ for intra-and inter-module exploration tendency.} For user groups in different ranges of $R_{g_c}$, for simplicity, $\rho_w \approx 0.6 $  and $\rho_c 
 \approx 1$.}
\label{SI_rho}
\end{figure*}

\begin{figure*}[t!]
\centering
\includegraphics[scale=1.0]{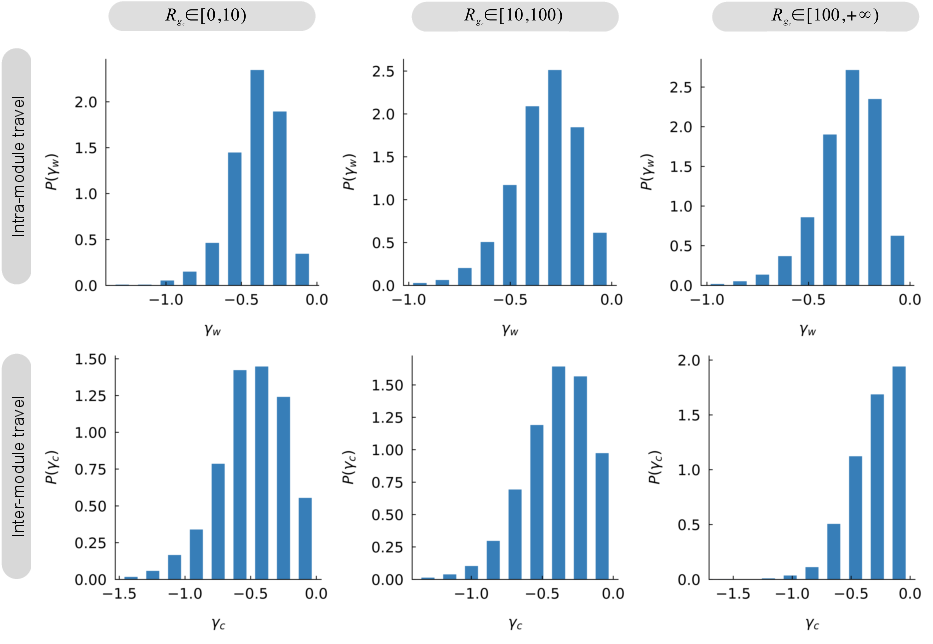}
\caption{{\bf The parameter $\gamma_w$ and $\gamma_c$ for intra-and inter-module exploration tendency.} For user groups in different ranges of $R_{g_c}$, for simplicity, $\gamma_w \approx 0.21 $  and $ \gamma_c$ increase with $R_{g_c}$.}
\label{SI_gamma}
\end{figure*}

We introduce the switch mechanism to the the standard exploration/preferential return scheme\cite{song2010modelling} to enable individuals to travel across modules, see Fig. \ref{SI_Model}. The model parameters $\rho_w$ ($\rho_c$) and $\gamma_w$ ($\gamma_c$) control users' tendency to explore new locations (new modules) are illustrated in Fig. \ref{SI_rho} and Fig. \ref{SI_gamma}. The mean value for $\rho_w$ and $\rho_c$ keep around 0.6 and 1 respectively. Same with conclusions in Fig. \ref{SI_Senegal_MSD} and Fig.\ref{SI_US_MSD}, the $\gamma_w$ keeps around 0.21, whereas $\gamma_c$ decrease for user groups in larger $R_{g_c}$. Table \ref{tableS1} summarizes the parameter settings used for the simulation.

\subsection*{3.2 Sensitivity analysis} 
\addcontentsline{toc}{subsection}{\protect\numberline{3.2} Sensitivity analysis}%

\begin{figure*}[t!]
\centering
\includegraphics[scale=0.9]{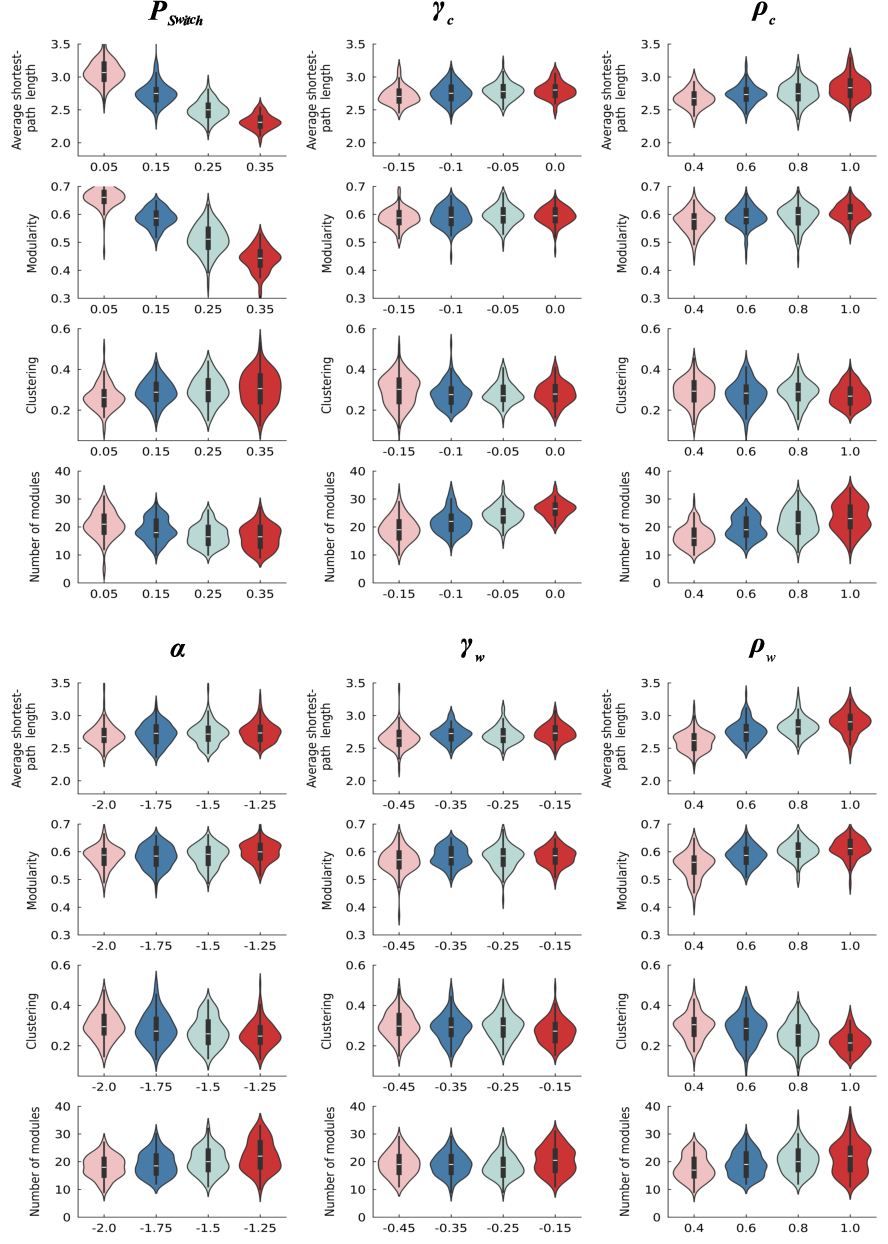}
\caption{{\bf Sensitivity analysis for the switch model.} By varying the model parameters—namely, the switch probability ($P_{Switch}$), inter-module exploration tendency ($\gamma_c$, $\rho_c$), the intra-module tendency ($\gamma_w$,$\rho_w$), and travel distance distribution ($\alpha$)—we evaluate their impact on the mobility in terms of their network characteristics.}
\label{SI_Sensitivity}
\end{figure*}

Multiple parameters in the proposed switch model will affect the outcomes of mobility networks and their characteristics. We use the parameters in Table \ref{tableS1} as the default settings. For each simulation, we vary one parameter at a time, running the switch model for 100 iterations and 1000 steps per individual. The designed experiments and the results are as follows (also see Fig. \ref{SI_Sensitivity}),
\begin{itemize}
 \item[1.] $P_{Switch}$ for the probability of switching to inter-module travel. By testing $\beta \in \{0.05,0.15,0.25,0.35\}$ and we can observe that larger $P_{Switch}$ leads to smaller shortest-path length and smaller modularity, but higher clustering.

 \item[2.] $\rho_c$ and $\gamma_c$ for inter-module exploration $P_c(R_{g_c})=\rho_c S_c^{\gamma_c}=\rho_c S_c^{-(b+a log(R_{g_c}))}$. We set $b=0.6$ and test $\rho_c \in \{0.4,0.6,0.8,1.0\}$ and  $a \in \{-0.15,-0.1, -0.05, -0\}$. We find that larger $\rho_c$ and larger $a$ lead to larger shortest-path length and larger modularity, but smaller clustering.

\item[3.] $\rho_w$ and $\gamma_w$ for intra-module exploration $P_w(R_{g_c})=\rho_w S_w^{-\gamma_w}$. We test $\rho_w \in \{0.4,0.6,0.8,1.0\}$ and  $\gamma_w \in \{0.15,0.25,0.35,0.45\}$. $\gamma_w$ make nearly no difference in model outcomes. Larger $\rho_w$ leads to larger shortest-path length and larger modularity, but smaller clustering.
 
\item[4.] $\alpha$ for jump-distance distribution $P(\Delta r) \sim \Delta r^{-1-\alpha}$. By testing $-(1+\alpha) \in \{-2.0,-1.75,-1.5,-1.25\}$, we can observe that larger $-(1+\alpha)$ leads to larger shortest-path length and larger modularity, but smaller clustering.

\end{itemize}
In short, increasing parameters $\rho_c$, $\rho_w$, and $\alpha$ primarily increase modularity. Increasing $P_{Switch}$ decreases modularity (When $P_{Switch}=0$, the model is equivalent to the EPR model, which produces low modularity but high clustering). It's worth mentioning that the parameters $\gamma_w$ and $\gamma_c$ have no apparent impact.

\subsection*{3.3 Predict mobility fluxes} 
\addcontentsline{toc}{subsection}{\protect\numberline{3.3} Predict mobility fluxes}%
 Besides reproducing trajectory networks, our model is also capable of predicting human movement at the collective level. We first summarize the fluxes/unique users between counties from massive trajectories and average the data from all iterations. The predictive ability of the model is measured by the Sorensen–Dice Similarity Index (SSI)\cite{dice1945measures}, that is,
\begin{equation}
\text{SSI}=\frac{1}{N}\sum_{\{i,j\} }\frac{2\text{min}(T_{ij},\bar{T}_{ij})}{T_{ij}+\bar{T}_{ij}}
\end{equation}
where $T_{ij}$ is the simulated fluxes/unique users between county $i$ and county $j$ and $\bar{T}_{ij}$ is the corresponding empirical fluxes/unique users between county $i$ and county $j$. A similarity of 1 means that the simulated data and empirical data are a perfect match.

Take U.S. data as an example. The total fluxes and the number of unique visitations between counties generated by our model agree well with the empirical data with SSI $>0.61$ in all groups (Extended Data Fig. 2). Notwithstanding that the EPR model performs well for populations in the medium range of $R_{g_c} \in [10,100)$km, the EPR model has lower SSIs for the population in the range of $R_{g_c} \in [0,10)$km and $R_{g_c}\in [100, \infty)$km. The results suggest that the EPR model mainly focuses on the travels of majority populations with $R_{g_c} \in [0,10)$km and ignores the travels of populations with $R_{g_c}\in [100, \infty)$km.

\clearpage
\bibliographystyle{Nature}
\bibliography{bib}
\clearpage

\begin{table}[!htbp]
\centering
\caption{Model parameters for U.S. data and Senegal data.}
\footnotesize
\begin{tabular}{c|c|c|c|c|c|c|c}
\hline
Parameters & $\alpha$ [$P(\Delta r)$] & $\beta$ [$P(\Delta t)$] & $\eta$ [$P({R_{g_c}})$] &($\rho_w$,$\gamma_w$) [$P_w$] & ($\rho_c$,$\gamma_c$) [$P_c$] &$P_{Switch}$ & {Module size}\\
\hline
U.S. data &0.50 &0.60 &0.12 & 0.6, 0.21 & 1, 0.58-0.12 $\log(R_{g_c})$ & 0.14 &10\\
\hline
Senegal data &1.26 &0.22 & 0.27 & 0.6, 0.21& 1, 1.06-0.30 $\log(R_{g_c})$ & 0.14 &5\\
\hline
\end{tabular}
Note: The arithmetic expression inside ``[ ]'' is to explain how the parameters are formulated. Module size means the number of unique locations within module.
\label{tableS1}
\end{table}
\clearpage

\begin{table}[!h]
\caption{Notations of the variables and parameters in this paper.}
\centering
\footnotesize
\begin{tabular}{p{2cm}p{8cm}p{4.5cm}}
\hline
Notations & Description & Related formulas \\
\hline
$T$ & Individual trajectory & $T=\{\theta_1,...,\theta_i,... \}$ \\
$G(T)$ & Individual trajectory network & $G(T)=\{T,W(T)\}$ \\
$W(T)$ & Weight set in  trajectory network & $W(T)=\{w(\theta_i,\theta_{i+1})\}$\\
$w(\theta_i,\theta_{i+1})$ & Weight between stay points &$w(\theta_{i},\theta_{i+1})=\log(\frac{\hat{d}}{d(\theta_{i},\theta_{i+1})})$\\
$d(\theta_i,\theta_{i+1})$ & Distance between stay points  &$w(\theta_{i},\theta_{i+1})=\log(\frac{\hat{d}}{d(\theta_{i},\theta_{i+1})})$\\
\hline
$Q$ & Network modularity \\
$C$ & Network average clustering coefficient\\
$L$ & Network average shortest-path length\\
\hline
$d_c$ & Distance of module from home & \\
$R_{g_c}$ & Inter-module radii of gyration & $P(R_{g_c}) \sim R_{g_c}^{-\eta}$ \\
$\Delta r$ & Jump distance & $P(\Delta r) \sim \Delta r^{-(1+\alpha)}$\\
$\Delta t$ & Stay time & $P(\Delta t) \sim \Delta t^{-(1+\beta)}$\\
$MSD(t)$ & Mean square displacement & $\text{MSD}(t)^{1/2} \sim \log(t)^{v}$\\
$P_w$ & Within-(Intra-) module exploration tendency & $P_w\sim \rho_w S_w^{-\gamma_w}$\\ 
$S_w$ & Number of uniquely visited locations within module & $P_w\sim \rho_w S_w^{-\gamma_w}$\\ 
$\rho_w$, $\gamma_w$ & Parameters for intra-module exploration tendency & $P_w\sim \rho_w S_w^{-\gamma_w}$ \\
$P_c$  & Cross-(Inter-) module exploration tendency & $P_c\sim \rho_c S_c^{-\gamma_c}$\\ 
$S_c$ & Number of uniquely visited modules & $P_c\sim \rho_c S_c^{-\gamma_c}$\\ 
$\rho_c$, $\gamma_c$ & Parameters for inter-module exploration tendency & $P_c\sim \rho_c S_c^{-\gamma_c}$\\ 
$P_{Switch}$ & Tendency of switch of exploration mode & \\
\hline
\end{tabular}
\label{table_notation}
\end{table}
\clearpage

\clearpage

\clearpage

\clearpage

\clearpage

\clearpage

\clearpage

\clearpage

\clearpage

\clearpage

